\documentclass[doublecol]{epl2} 
\usepackage{psfrag}
\usepackage{amssymb}

\title{First-principles quantum dynamics for fermions: \\Application to
molecular dissociation}
\shorttitle{First-principles quantum dynamics for fermions} %Insert here a short version of the title if it exceeds 70 characters

\author{Magnus \"{O}gren K.~V.~Kheruntsyan and J.~F.~Corney}
\shortauthor{M. \"{O}gren \etal}

\institute{                    
ARC Centre of Excellence for Quantum-Atom Optics, School of Mathematics and
Physics, University of Queensland, Brisbane, Queensland 4072, Australia\\
}

\abstract{We demonstrate that the quantum dynamics of a many-body Fermi-Bose system
can be simulated using a Gaussian phase-space representation
method. In particular, we consider the application of the mixed fermion-boson model
to ultracold quantum gases and simulate the dynamics of dissociation of a
Bose-Einstein condensate of bosonic dimers into pairs of fermionic atoms. We
quantify deviations of atom-atom pair correlations from Wick's factorization
scheme, and show that atom-molecule and molecule-molecule correlations grow
with time, in clear departures from pairing mean-field
theories. As a first-principles approach, the method provides benchmarking
of approximate approaches and can be used to validate dynamical probes for
characterizing strongly correlated phases of fermionic systems.
}

\begin{document}

\maketitle

\section{Introduction}

The physics of interacting fermions is the basis of many of the most
important phenomena in condensed matter physics, ultracold gases, and
quantum chemistry. A fundamental issue is how the microscopic interactions
at the quantum level give rise to collective and emergent effects in
many-body systems. For many situations, particularly in condensed matter
systems, static or equilibrium correlation functions are sufficient to
connect theory and experiment, and sophisticated techniques have been
developed to calculate and measure them \cite{Vojta2003}.

Addressing similar questions in the domain of many-body \emph{dynamics}, however, has limitations in condensed matter systems. Ultracold quantum
gases, on the other hand,
allow creation
of highly controllable implementations of analogue many-body
systems for which the dynamical evolution and correlations are directly
accessible \cite{Bloch2002,DurrPRA2004,GreinerPRL2005,Esslinger2005,Westbrook2007}. The
purity and tunability of these `tailor-made' analogue systems means
that ultracold quantum gases are ideal for testing fundamental ideas in
quantum many-body physics and are leading candidates for dynamical `quantum
simulation' \cite{Bloch2008,Rigol,CazalillaRigol,Buluta2009}. 
In order to make predictions from the underlying theory and to
validate the potential simulators~\cite{Troyer,CazalillaRigol}, or to benchmark approximate
approaches, a numerical simulation of the 
real-time dynamics is
required. Similar requirements of exact simulation of many fermions
arise in determining the quantum chemistry of complex molecular systems~\cite%
{Miller2005}.

In this work we perform first-principles dynamical simulations of a
fermion-boson model.  We use a Gaussian stochastic method based on a
generalized phase-space representation of the quantum density operator~\cite%
{CorneyPRL2004}. The fermion-boson model forms the underlying basis for a
broad range of phenomena in condensed matter and ultracold atom physics. It
was originally proposed in the context of high-temperature superconductivity~%
\cite{FriedbergPRB1989}, but in ultracold gases it corresponds to the theory
of resonance superfluidity with Feshbach molecules~\cite{Feshbach}. The
latter forms the basis of a two-channel model for describing the physics of
the BCS-BEC crossover~\cite{Griffin}. More recently, the fermion-boson model
has been used for analyzing the decay of double occupancies 
(doublons)~\cite{Altman} in a driven 
Fermi-Hubbard system \cite{Esslinger2008}. The particular situation that
we simulate here corresponds to spontaneous dissociation of a Bose-Einstein
condensate (BEC) of molecular dimers into fermionic atoms~\cite%
{JackPRA2005,KheruntsyanPRL2006}, in which case the model provides the
fermionic equivalent of parametric downconversion in quantum optics: the
production of pairs of entangled particles.

The Gaussian phase-space method represents an extension to fermions of  successful bosonic techniques
\cite{positiveP,GardinerBook1,Drummond-Carter-97,Steel1998,GilchristPRA1997,Deuar2002,Corney2003,Savage2006,Deuar2007,Perrin2008}.
The essence of the method is the mapping of the density operator evolution onto a Fokker-Planck
equation for a phase-space distribution, via a continuous Gaussian operator
basis~\cite{CorneyPRL2004}. The evolving distribution is then sampled with
stochastic differential equations (SDEs) for the phase-space variables~\cite%
{positiveP,GardinerBook1}. The mapping to the phase-space distribution is 
exact \cite{positiveP,GardinerBook1,GilchristPRA1997,CorbozPRB2008} provided no boundary terms arise in deriving the
the Fokker-Planck equation. In practice, such terms may develop after some simulation time, but in an easily verifiable way \cite{GilchristPRA1997,CorbozPRB2008}, putting a well characterised upper limit to a useful simulation duration.    Numerical signatures of systematic errors include:  (\textit{i}) the onset of spiking behaviour due to the presence of  near-singular trajectories; (\textit{ii}) sudden dramatic increase in the statistical uncertainties; and (\textit{iii}) development of power-law tails in the probability distribution.
All these signatures, well-known from the early studies of real-time dynamics of bosonic systems \cite{GilchristPRA1997} and from equilibrium calculations for fermion systems using imaginary-time techniques \cite{CorneyPRL2004,CorbozPRB2008}, carry over to the present simulations of fermion dynamics and are verified in the numerical examples that we present below.

As in any stochastic method, sampling error limits the precision of the results.  However, unless the distribution develops power-law tails, indicated by the above-mentioned signatures, this uncertainty can be made arbitrarily small by increasing the number of trajectories.

From the physical point of view, the Gaussian phase-space method can be viewed as providing the quantum
corrections, through additional stochastic terms, to different mean-field
approaches. For example, with certain factorization assumptions~\cite%
{RahavPRB2009}, the method reduces to a time-dependent
Hartree-Fock formalism. %(TDHF).
Furthermore, neglecting the stochastic terms recovers the approximate
pairing mean-field theory (PMFT)~\cite{JackPRA2005,DavisPRA2008}, to which
we compare the phase-space results. While often accurate for determining particle
number densities, the mean-field approach gives no direct information about
higher-order correlations, and its accuracy is not known \textit{a priori}.
In contrast to this, the first-principles simulations presented here reveal 
significant development of higher-order correlations.

For the first application of the fermionic phase-space method to a multimode
dynamical problem, we consider a uniform molecular BEC (MBEC) initially in a
coherent state at zero temperature, with no atoms present.
Assuming sufficiently low densities, we neglect $s$-wave scattering
interactions to simplify the treatment. The Hamiltonian of this
fermion-boson model~\cite{FriedbergPRB1989} is given by
\begin{equation}
\widehat{H}=\hbar \sum\nolimits_{\mathbf{k},\sigma }\Delta _{\mathbf{k}}\hat{%
n}_{\mathbf{k},\sigma }-i\hbar \kappa \sum\nolimits_{\mathbf{k}}\left( \hat{a%
}^{\dagger }\hat{m}_{\mathbf{k}}-\hat{m}_{\mathbf{k}}^{\dagger }\hat{a}%
\right),  \label{Hamiltonian}
\end{equation}%
where $\mathbf{k}$ labels the plane-wave modes and $\sigma =1,2$ labels the
effective spin state for the atoms. Even though we will present the
numerical results for a one-dimensional (1D) system, we formulate the
problem in the general case as the method is straightforward to use in
higher dimensions. The fermionic number and pair operators are defined as $\hat{n}_{\mathbf{k}%
,\sigma }\!=\!\hat{c}_{\mathbf{k},\sigma }^{\dagger }\hat{c}_{\mathbf{k},\sigma }
$ and $\hat{m}_{\mathbf{k}}\!=\!\hat{c}_{\mathbf{k},1}\hat{c}_{-\mathbf{k},2}$,
with $\{\hat{c}_{\mathbf{k},\sigma },\hat{c}_{\mathbf{k}^{\prime },\sigma
^{\prime }}^{\dagger }\}\!\!=\!\!\delta_{\mathbf{k}\mathbf{k}^{\prime }}\delta
_{\sigma \sigma ^{\prime }}$,
while the bosonic molecular operator obeys $[\hat{a},\hat{a}^{\dagger }]\!=\!1$.
The atom-molecule coupling (invoked by a magnetic Feshbach resonance
sweep or optical Raman transitions) is characterized by $\kappa
=\chi_{D}/L^{D/2}$ \cite{DavisPRA2008}, where $L$ is the size of the
quantization box, and mediates an effective interaction between the atoms. The first term, $\hbar\Delta _{\mathbf{k}}\!\equiv \!\hbar^{2}
\left\vert \mathbf{k}\right\vert ^{2}/(2m_{a}) + \hbar\Delta$, contains the
kinetic energy of the atoms (of mass $m_{a}$), while the detuning $\Delta<0$
corresponds to the total dissociation energy $2\hbar \left\vert \Delta
\right\vert$ imparted onto the system by the external fields.

Because of the symmetry between spins in the Hamiltonian, and the equal
initial populations, we need only to consider the number operator for one of the spin states $\hat{n}_{\mathbf{k}}=\hat{n}%
_{-\mathbf{k}}=\hat{n}_{\mathbf{k},1}=\hat{n}_{\mathbf{k},2}$. An additional
operator identity that follows from the Hamiltonian is
\begin{equation}
\hat{m}_{\mathbf{k}}^{\dagger }\hat{m}_{\mathbf{k}} \left( =\hat{n}_{\mathbf{%
k},1}\hat{n}_{-\mathbf{k},2} \right) =\hat{n}_{\mathbf{k}},
\label{MPMrelation}
\end{equation}
which arises because the condensate to which the atom pairs are coupled is
assumed to be homogeneous. One consequence of eq.~(\ref{MPMrelation}) is
that the relative number of atoms with equal and opposite momenta is
perfectly squeezed~\cite{KheruntsyanPRL2006}, i.e. with zero variance. It
also means that the second-order atom-atom correlation function reduces to $%
g_{12}^{(2)}(\mathbf{k},-\mathbf{k}) \equiv \langle \hat{m}_{\mathbf{k}%
}^{\dagger }\hat{m}_{\mathbf{k}}\rangle /\langle \hat{n}_{\mathbf{k}%
,1}\rangle \langle \hat{n}_{-\mathbf{k},2}\rangle =1/\langle \hat{n}_{%
\mathbf{k}}\rangle $. Thus the atom-atom correlation function can be
determined from the number density alone.

One effective approximate approach for treating the dynamics of dissociation
is the PMFT~\cite{JackPRA2005,DavisPRA2008}, which is obtained by assuming
atom-molecule decorrelation and by replacing the molecular operator by a
coherent mean-field amplitude, $\hat{a} \rightarrow \beta$.

In this paper we solve the full Hamiltonian (\ref{Hamiltonian}) exactly, and
in order to quantify deviations from the PMFT behavior we evaluate several
correlation functions. The departures from Wick decorrelation are
analyzed via the correlation coefficient
\begin{equation}
W=\sum\nolimits_{\mathbf{k}}\langle \hat{m}_{\mathbf{k}}^{\dagger }\hat{m}_{%
\mathbf{k}}\rangle \,/\,\sum\nolimits_{\mathbf{k}}\left( \left\vert \langle
\hat{m}_{\mathbf{k}}\rangle \right\vert ^{2}+\langle \hat{n}_{\mathbf{k}%
}\rangle ^{2}\right),  \label{WTPMFT}
\end{equation}%
which is unity within the PMFT.

To examine molecule-atom pair correlations and the second-order coherence of
the molecular field, we define
\begin{equation}
g_{ma}^{(2)}(\mathbf{k})=\frac{\langle \hat{a}^{\dagger}\hat{a}\hat{n}_{%
\mathbf{k}}\rangle }{\langle \hat{a}^{\dagger}\hat{a}\rangle \langle \hat{n}%
_{\mathbf{k}}\rangle },\,\, g_{mm}^{(2)} =\frac{\langle \hat{a}^{\dagger}%
\hat{a}^{\dagger}\hat{a}\hat{a}\rangle }{\langle \hat{a}^{\dagger}\hat{a}%
\rangle ^{2}}.  \label{g-at-mol}
\end{equation}
Again, within the PMFT, these will be unity. We may expect that, over time,
correlations will develop between the molecular and atomic fields; the Gaussian phase-space 
simulations give exact quantitative accounts of these
effects.

\section{Fermionic phase-space representation}
As mentioned above, the essence of the Gaussian phase-space method is the mapping 
of the density operator evolution into a set of stochastic differential equations \cite{CorneyPRL2004} 
that can be solved on a computer. For a large class of 
second-quantized system Hamiltonians, 
such as those containing no higher than quartic terms in field operators, such mapping introduces 
no additional approximations. 
 For $M$ plane-wave modes, the quantum state governed by Hamiltonian (\ref%
{Hamiltonian}) can be mapped onto a complex phase-space of dimension $3M+2$: $\vec{\lambda}(t)=(n_{1},\dots
,n_{M},m_{1},\dots ,m_{M},m_{1}^{+},\dots ,m_{M}^{+},\beta ,\beta ^{+}) $,
with $m_{j}^{+}\neq m_{j}^{\ast }$ and $\beta^{+}\neq \beta^{\ast }$. The phase-space equations are structurally similar to the Heisenberg
equations for the corresponding operators, but the stochastic terms are not unique and can be modified by different choices of diffusion gauge \cite{Deuar2002,CorneyPRL2004,Ogren2010}. This freedom allows the equations to be tailored to have different numerical
properties. One specific set of Ito SDEs is \cite{Ogren2010}:
\begin{equation}
\begin{array}{l}
\dot{n}_{\mathbf{k}}=\alpha m_{\mathbf{k}}^{+}+\alpha ^{+}m_{\mathbf{k}%
}+ \frac{1}{\sqrt{N_{0}}} n_{\mathbf{k}}\left( m_{\mathbf{k}}\zeta _{1}^{\ast }+m_{%
\mathbf{k}}^{+}\zeta _{2}^{\ast }\right) \\
\dot{m}_{\mathbf{k}}=-2i\delta _{\mathbf{k}}m_{\mathbf{k}}+\alpha \left(
1-2n_{\mathbf{k}}\right) + \frac{1}{\sqrt{N_{0}}} \left( m_{\mathbf{k}}^{2}\zeta
_{1}^{\ast }-n_{\mathbf{k}}^{2}\zeta _{2}^{\ast }\right) \\
%\ \ + \frac{1}{\sqrt{N_{0}}} \left( m_{\mathbf{k}}^{2}\zeta
%_{1}^{\ast }-n_{\mathbf{k}}^{2}\zeta _{2}^{\ast }\right) \\
\dot{m}_{\mathbf{k}}^{+}=2i\delta _{\mathbf{k}}m_{\mathbf{k}}^{+}+\alpha
^{+}\left( 1-2n_{\mathbf{k}}\right) + \frac{1}{\sqrt{N_{0}}} \left( m_{\mathbf{k}%
}^{+2}\zeta _{2}^{\ast }-n_{\mathbf{k}}^{2}\zeta _{1}^{\ast }\right)\\
%\ \ + \frac{1}{\sqrt{N_{0}}} \left( m_{\mathbf{k}%
%}^{+2}\zeta _{2}^{\ast }-n_{\mathbf{k}}^{2}\zeta _{1}^{\ast }\right) \\
\dot{\alpha}=-\frac{1}{N_{0}}\sum_{\mathbf{k}}m_{\mathbf{k}%
}+ \frac{1}{\sqrt{N_{0}}} \zeta _{1} \\
\dot{\alpha}^{+}=-\frac{1}{N_{0}}\sum_{\mathbf{k}}m_{\mathbf{k}%
}^{+}+ \frac{1}{\sqrt{N_{0}}} \zeta _{2},%
\end{array}
\label{SDEexample}
\end{equation}%
where the derivative is with respect to a scaled time, $\tau =t/t_{0}$, with
$t_{0}=1/\kappa \sqrt{N_{0}}$. We have normalized the molecular field by its
maximum (initial) value, $\alpha=\beta/\sqrt{N_{0}}$, where $N_{0}$ is the
initial number of molecules. The 
complex Gaussian noises $\zeta _{j}$ obey $\langle \zeta _{j}(\tau ) \zeta
_{j^{\prime }}(\tau ^{\prime }) \rangle =0,\,\langle \zeta _{j}(\tau ) \zeta
_{j^{\prime }}^{\ast } (\tau ^{\prime }) \rangle =\delta _{jj^{\prime
}}\delta (\tau -\tau ^{\prime })$. This form of eqs.~(\ref{SDEexample})
shows that with drift terms of order $1$, the
noise terms are $\sim 1/\sqrt{N_{0}}$, i.e. the noise and therefore
non-mean-field corrections to correlations become more important for
decreasing $N_{0}$. In practice we convert the equations to Stratonovich
form and integrate with a semi-implicit method. Stochastic averages of the
variables give the first-order operator moments; normally ordered
higher-order moments are obtained by averages of the corresponding Wick
decomposition~\cite{CorneyPRL2004}, e.g. $\langle \hat{m}_{\mathbf{k}%
}^{\dagger }\hat{m}_{\mathbf{k}}\rangle =\langle m_{\mathbf{k}}^{+}m_{%
\mathbf{k}}\rangle _{S}+\langle n_{\mathbf{k} }^{2}\rangle _{S}$.
Note that the final, averaged moment will not satisfy Wick's theorem for a
general quantum state.

The stochastic sampling assumes a sufficiently bounded distributions, such that any boundary terms could have been neglected in obtaining
the Fokker-Planck equation. Previous experience with bosons~\cite%
{GilchristPRA1997} and fermions~\cite{CorbozPRB2008} has shown that spikes and a sudden rapid growth of the statistical sampling errors
in observables are seen when the tails of the probability distributions do not decay fast enough. Through comparisons with independent numerical solutions we 
confirm that the onset of such spiking behaviour and the rapidly growing sampling error signifies the limit of the useful simulation time. Although it can be controlled somewhat with gauges, \cite{Deuar2002,CorneyPRL2004,Ogren2010}, the finite simulation time is a well-known limitation of
stochastic phase-space methods. 
In the numerical examples of figs. \ref{Fig1}, \ref{Fig3} and \ref{Fig4} below, the simulation results are shown for time windows well below the spiking time. We simulate a sufficiently large number of stochastic trajectories to reduce the statistical errors to below the thickness of the relevant curves shown in the figures. In addition, we use identities such as (\ref{MPMrelation}) and different gauges as a further demonstration that the
simulations are exact up until the emergence of spiking behaviour. 

Unlike quantum Monte Carlo approaches that are well-suited to calculation of
exact ground-state properties or to simulation through imaginary time~\cite%
{Linden1992}, the Gaussian stochastic method does not suffer from a
`dynamical sign problem'~\cite{Mak1994}. Other approaches for
real-time simulations include the time-dependent density functional theory~\cite{RungePRL1984}, although in practice this method is often
restricted in accuracy by the need for exact functionals. Methods that use
matrix-product-states based algorithms have been very successful for
applications to one spatial dimension~\cite%
{CazalillaPRL2002,DaleyVidal2004,White2004,Giamarchi2006,Cirac2009},
however, as these methods require a truncated basis they do not fulfill the
strict benchmarking criteria that a first-principles method can provide.
An interesting direction in recent
years has been the extension to fermionic systems of stochastic
wavefunction approaches~\cite{MontinaPRA2006}, which are similar
in spirit to phase-space methods.

\begin{figure}[tbh]
\psfrag{x}{$\tau$}  %%% Karen's label fix.
\psfrag{t}{$\tau$}
\psfrag{y}{$\langle \hat{n}_{k}\rangle $}
\psfrag{N}{$N_{a},N_{m}$} %
\includegraphics[width=8cm]{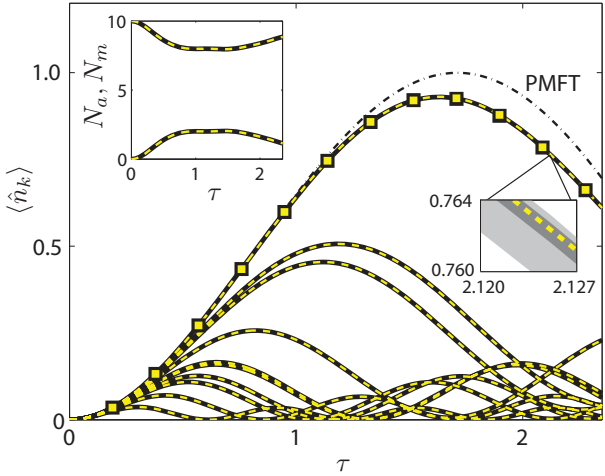}
\caption{(Color online) Comparison between different methods for the population dynamics of individual atomic modes $\langle \hat{n}_{k}(\tau)\rangle $: phase-space
method with $10^6$ stochastic trajectories (solid black curves), number-state basis (dashed yellow curves). The top curve with the largest oscillation amplitude is for the
resonant mode $k_0=6dk$ (corresponding to $\protect\delta_{k_{0}}=t_{0}%
\Delta_{k_{0}}=0$); the other curves are the sidebands stepped by $%
dk=160^{-1/4}/d_{0}$, where the lengthscale is $d_{0}=\protect\sqrt{\hbar
t_{0}/2m_{a}}$. For the $k_{0}$-mode, we also plot the result from the PMFT
for comparison (top dash-dotted curve). To illustrate the identity (\protect
\ref{MPMrelation}) we plot the quantity $\langle \hat{m}_{k_{0}}^{\dagger }\hat{m}%
_{k_{0}}\rangle $ as large black squares for the phase-space method and as small
yellow squares for the number-state calculation. The left inset shows the number of molecules $N_m=\langle \hat{a}^{\dagger}\hat{a}\rangle$ (top curve) and the total number of atoms $N_a= \sum_k
\langle \hat{n}_{k} \rangle$ in one of the spin states.
For all curves from the phase-space method, the statistical errors of $\pm 1\sigma$ standard deviation are smaller than the thickness of the curves, within the time window shown. The
inset on the right shows these errors explicitly as dark and light grey shadings corresponding to quantities $\langle \hat{n}_{k_{0}}(\tau)\rangle $ and $\langle \hat{m}_{k_{0}}^{\dagger }\hat{m}_{k_{0}}\rangle $, respectively.}
\label{Fig1}
\end{figure}

\section{Few-mode system}
To confirm the validity of our numerical implementation of the
phase-space method, we first independently solve a small system with $N_{0}=10$
molecules and $M=10$ atomic modes in a standard number-state basis. For this
test system, with a bosonic number-basis truncation of $n_{\mathrm{max}}\sim
10^{2}$, the Hilbert space has dimension $%
d=2^{M}n_{\mathrm{max}}\simeq 10^{5}$. In fig.~\ref{Fig1} we show the
population in the momentum modes $\langle\hat{n}_{k}\rangle$ calculated
using the phase-space method (with $10^6$ stochastic trajectories) and the number-state basis; we also illustrate
the identity (\ref{MPMrelation}) by calculating and plotting $\langle \hat{m}%
^{\dagger }_{k_{0}} \hat{m}_{k_{0}} \rangle $ directly and comparing it with
$\langle\hat{n}_{k_{0}}\rangle$. The agreement between the two methods is
excellent. The top two curves in fig.~\ref{Fig1} illustrate the deviation of
the PMFT prediction (dashed-dotted curve) from the exact calculation (solid
curve) for the resonant mode $k_0$.

\begin{figure}[tbp]
\includegraphics[width=7cm]{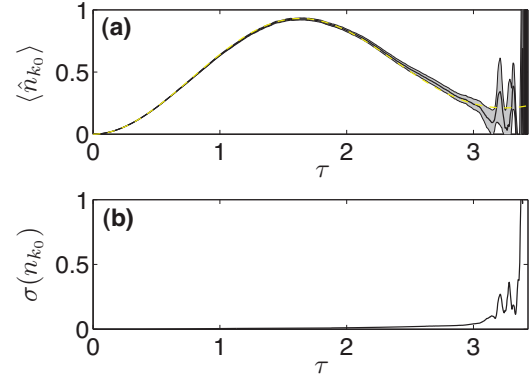}
\caption{(Color online) (a) Population of the resonant mode, $\langle \hat n_{k_0}\rangle $, as a function of time $\tau$ for the same parameters as in fig. \ref{Fig1}, except with $10^3$ stochastic trajectories and a longer simulation time. The solid line is from the phase-space method, with the grey shading representing statistical errors of $\pm1\sigma$ standard deviation. The dashed yellow line is from the number-state calculation. (b) Evolution of the statistical error on $n_{k_0}$, showing the emergence of spiking behaviour past $\tau \gtrsim 3$ and the subsequent rapid growths of the sampling errors. The size of the statistical errors -- prior to the emergence of the spiking behaviour -- scales as $1/\sqrt{N_s}$ as expected from the central limit theorem, where $N_s$ is the number of stochastic trajectories.
}
\label{Fig2}
\end{figure}

In fig. \ref{Fig2}, we show the population dynamics of the resonant mode, $n_{k_0}$, together with the explicit behaviour of the statistical errors due to the stochastic sampling. The number of stochastic trajectories in this example (which otherwise is the same as in fig. \ref{Fig1}) is small enough to render the sampling errors visible. At the same time, we have chosen the time window longer than the onset of spiking behaviour and the sudden dramatic growth of statistical errors to explicitly illustrate the signatures of systematic errors that limit the useful simulation time of the phase-space method.

\begin{figure}[tbp]
\includegraphics[width=8.5cm]{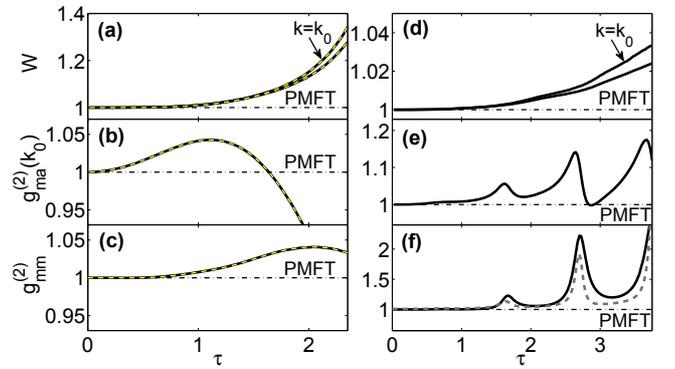}
\caption{(Color online) (a), (b), (c) Correlation coefficients $W$, $%
g_{ma}^{\left(2\right)}(k_{0})$, and $g_{mm}^{\left(2\right)}$ as a function
of time for the test system of fig.~\protect\ref{Fig1}, with $N_{0}=10$ and $%
M=10$. Solid black curves are from the phase-space method; dashed
yellow curves are from the number-state calculation.
The lower curve in
(a) is for the full summation in eq.~(\protect\ref{WTPMFT}), whereas the
upper curve is the respective correlation coefficient for the resonant mode $%
k_{0}$. (d), (e), (f) The same as in the left column, but from the phase-space
method for $N_{0}=10^{2}$ and $M=10^{3}$. The dashed grey curve in (f) is
from an ensemble of PMFT calculations with a set of Poissonian-weighted $N_{0,j}$ and $
\overline{N}_{0} =10^{2}$ (see text). Statistical errors of $\pm 1\sigma$ standard deviation for all curves from the phase-space method are contained within the thickness of the curves and are obtained from $10^5$ stochastic trajectories.}
\label{Fig3}
\end{figure}

To further evaluate the differences between treating the Hamiltonian (\ref%
{Hamiltonian}) exactly and using the approximate PMFT, we plot in fig.~\ref%
{Fig3}~(a) the correlation coefficient $W$. Clear deviations are seen as
time evolves; the deviations illustrate that in the exact treatment the
following inequality holds, $\langle\hat{m}_{\mathbf{k}}^{\dagger}\hat{m}_{%
\mathbf{k}}\rangle\geq\left|\left\langle \hat{m}_{\mathbf{k}}\right\rangle
\right|^{2}+\left\langle \hat{n}_{\mathbf{k}}\right\rangle ^{2}$, whereas
the PMFT prescribes an equality sign. 
Next, we consider the molecule-atom and molecule-molecule second-order
correlations, $g_{ma}^{(2)}$ and $g_{mm}^{(2)}$ [see figs.~\ref{Fig3} (b)
and (c)]. Within the PMFT, both correlations are identically equal to 1.
However, our exact results show that the molecule-atom correlation initially
grows with time while the total atomic population grows. 
Then it changes to anti-correlation as the resonant-mode atoms start to re-associate.
Meanwhile the effect on the molecular field from the atom interactions is revealed 
as it gradually loses its second-order coherence, albeit not by a
significant amount for this few-mode system.

\section{Multi-mode system}
\begin{figure}[tbp]
\includegraphics[width=7cm]{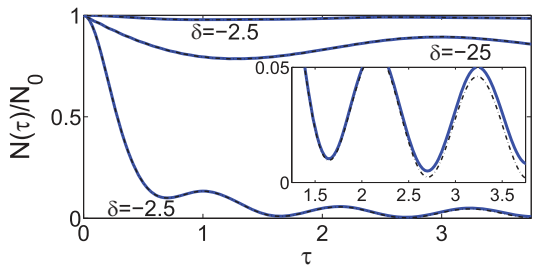}
\caption{(Color online) The fraction of remaining molecules, $N(\protect\tau)/N_{0}$, as a
function of time, for: $N_{0}=10^4$, $\protect\delta=-2.5$ (top curve); $%
N_{0}=10^2$, $\protect\delta=-25$ (intermediate curve); $N_{0}=10^2$, $%
\protect\delta=-2.5$ (bottom curve). In all cases the coupling $\protect%
\kappa$ is chosen to result in the same timescale $t_{0}=1/\protect\kappa
\protect\sqrt{N_0}=2\times 10^{-4}$ s. The solid curves are from the phase-space method (with $10^5$ trajectories), whereas the dash-dotted curves are from the approximate PMFT method; the difference between the
two types of curves (seen in the inset, for the lowest curve) is almost invisible on the scale of the graph. The curve for  $\protect\delta=-25$ illustrates a route away from the
regime of strong molecular depletion by increasing the dissociation energy $%
2\hbar\left|\Delta\right|$ by an order of magnitude, while keeping the same $%
N_{0}$ as for the lowest curve. For all curves from the phase-space method, the statistical errors of $\pm 1\sigma$ standard deviation are smaller than the thickness of the curves, within the time window shown.}
\label{Fig4}
\end{figure}

We now use the phase-space method for
simulating large 1D systems, with $M=10^{3}$ atomic modes and $%
N_{0}=10^{2}-10^{4}$ ($^{40}K_2$) molecules at densities $n_{1D}\simeq 1.3
\times 10^{5} -1.3 \times 10^{7}$ m$^{-1}$. In these cases, the number-state
calculation is impossible as the dimension of the Hilbert space is enormous (%
$d=2^{M} n_{\mathrm{max}}\gg10^{300}$). In fig.~\ref{Fig4} we show the
evolution of the number of molecules for three different cases. For the top
curve, the initial number ($N_{0}=10^{4}$) is much larger than the number of
available atomic modes, each of which hosts at most 1 atom due to the Pauli
blocking. Accordingly, we see negligible depletion of the MBEC, which makes
the relative size of the bosonic fluctuations very small. Hence, we do not
observe significant deviations from the PMFT, including in the molecular
second-order coherence, eq.~(\ref{g-at-mol}), which differed from 1 by less
than $10^{-5}$ in this case.

The situation changes for the bottom curve, for which $N_{0}=10^{2}$ is
comparable with the number of atomic modes within the relevant width of the
momentum distribution near $k_0$; we estimate this number~\cite{DavisPRA2008}
to be $\sim 0.1 M=10^{2}$. In this case, we see strong molecular depletion
and an increased role of bosonic quantum fluctuations so that the PMFT
starts to show disagreement with the exact result. Admittedly the
disagreement is still very small, implying that the predictions of the PMFT
for \emph{total particle numbers} can be rather accurate. The same is not
true, however, for higher-order correlations, shown in the right
column of fig.~\ref{Fig3} for the same parameters as the lowest curve of
fig.~\ref{Fig4}. Here, the large depletion of the MBEC and the increased role of
quantum fluctuations are manifest  -- beyond the predictability of the
PMFT -- in strong higher-order correlations. The correlation coefficient $W$
clearly deviates from one, though to a lesser extend than in the few-mode
system. The deviations of the molecule-atom and molecule-molecule correlations from 
$g_{ma}^{(2)}(k_{0})=1$ and $g_{mm}^{(2)}=1$, on the other hand, are
more dramatic.

The development of decoherence in the molecular field can
largely be accounted for by the dephasing of atomic-molecular oscillations due to total number uncertainty.  The frequency of the oscillations depends on the initial number of molecules; with a range of frequencies, the oscillations get out of phase and thus, for example, prevent complete disassociation of the molecular field from being seen in the average.  
As illustrated by the dashed grey curve in fig.~\ref{Fig3} (f), this effect can be reproduced by an ensemble of mean-field trajectories with different initial numbers.   
Here we have used a Poissonian weighting to give the same number-distribution as the initial coherent molecular field in the exact calculation, with $\overline
N_{0} =100$\cite{poisson}.   The large values of $g_{mm}^{(2)}$, which are possible with a state containing a superposition (or mixture) of a few low-occupation number states \cite{toystate}, occur at maximum depletion when the number-uncertainty is relatively large.

\section{Summary}
We have demonstrated a successful application of a fermionic
phase-space representation to first-principles quantum dynamics of a
fermion-boson model. We simulated the coherent molecular dissociation to
fermionic atoms and found significant higher-order correlations that cannot
be accounted for by the approximate pairing mean-field theory. The knowledge
of such correlations and the development of experimental probes to measure
them provide the most accurate characterization of quantum many-body phases
in strongly correlated systems.

The method is exact up until clear `spiking' signatures emerge in the stochastic trajectories. For the present model, the useful simulation duration encompasses the time required for partial recombination of the atoms into molecules and for significant higher-order correlations to emerge.
The accuracy 
during this useful simulation time was independently confirmed by comparison to number state calculations (for a small system) and by checking of conserved quantities.

Although we have here reported only on 1D simulations, we have also implemented 2D and 3D calculations and found that the method works reliably in higher dimensions. Extensions of the method to implement $s$-wave scattering interactions will
enable the study of non-equilibrium dynamics in a broader class of fermionic
systems of current experimental interest, such as atomic Mott insulators in
optical lattices and the BCS-BEC crossover problem.

\acknowledgments
The authors acknowledge support by the Australian Research Council.


\begin{thebibliography}{0}
\bibitem{poisson} More specifically, we perform an ensemble of PMFT 
calculations, each one having a different initial average number of molecules $N_{0,j}$, ranging over integer values between $N_{0,j}=50$ and $N_{0,j}=150$. The results of each simulation are then averaged with a Poissonian weight factor 
$w_j= (\overline{N}_{0})^{N_{0, j}}\exp( - \overline{N}_{0})/N_{0,j}!$.
The resulting pair correlation is given by  $g^{(2)}_{mm}= \sum_j w_j |\alpha_j|^4 / ( \sum_j w_j |\alpha_j|^2 )^2$.
\bibitem{toystate} As a simple example, consider the  state $| \Psi \rangle = (\sqrt{61} | 0 \rangle + | 1 \rangle + 3 \sqrt{2} | 2 \rangle )/4 \sqrt{5}$.  It has a value of $g^{(2)} \simeq 2.1$, corresponding to the peak in Fig. \ref{Fig3}~(f)  at $\tau \simeq 2.7$, where $\langle \hat{a}^\dagger \hat{a} \rangle \simeq 0.5$.
\end{thebibliography}
\end{document}